\documentclass[letter,traditabstract]{aa} 

\usepackage{graphicx}
\usepackage{txfonts}

\usepackage{epsf}
\usepackage{rotating}
\usepackage{graphics}

\def \sophie{{\it SOPHIE}}
\def \sophie{SOPHIE}
\def \MJ{M$_{\mathrm{Jup}}$}
\def \RJ{R$_{\mathrm{Jup}}$}
\def \RS{R$_{\odot}$}

\def \kms{km\,s$^{-1}$}
\def \ms{m\,s$^{-1}$}
\def \1s{$1\,\sigma$}
\def \kid{$\chi^2$}
\def \t0{T$_0$}

\def \cible{HAT-P-6}
\def \cibleb{{\cible}b}

\begin{document}

   \title{The retrograde orbit of the \cibleb\ exoplanet\thanks{Based 
   on observations collected with the 
              \sophie\  spectrograph on the 1.93-m telescope at
              Observatoire de Haute-Provence (CNRS), France, 
              by the \sophie\   Consortium (program 10A.PNP.CONS).
              \sophie\ radial velocities are available in electronic 
              form at the CDS via anonymous ftp to cdsarc.u-strasbg.fr 
              (130.79.128.5) or via 
              http://cdsweb.u-strasbg.fr/cgi-bin/qcat?J/A+A/.}}

   \author{G.~H\'ebrard\inst{1,2}, 
                 D.~Ehrenreich\inst{3},
	        F.~Bouchy\inst{1,2}, 
                 X.~Delfosse\inst{3},
                 C.~Moutou\inst{4}, 
                 L.~Arnold\inst{2},
                 I.~Boisse\inst{5,1},
                 X.~Bonfils\inst{3},
                 R.~F.~D\'{\i}az\inst{1,2},
                 A.~Eggenberger\inst{3},  
                 T.~Forveille\inst{3}, 
                 A.-M.~Lagrange\inst{3},
                 C.~Lovis\inst{6},
                 F.~Pepe\inst{6},
                 C.~Perrier\inst{3}, 
                 D.~Queloz\inst{6},
	        A.~Santerne\inst{4},
                 N.~C.~Santos\inst{5,7},
                 D.~S\'egransan\inst{6},
	        S.~Udry\inst{6}, 
                 A.~Vidal-Madjar\inst{1}
}

   \institute{
Institut d'Astrophysique de Paris, UMR7095 CNRS, Universit\'e Pierre \& Marie Curie, 
98bis boulevard Arago, 75014 Paris, France 
\email{hebrard@iap.fr}
\and
Observatoire de Haute-Provence, CNRS/OAMP, 04870 Saint-Michel-l'Observatoire, France
\and
UJF-Grenoble\,1\,/\,CNRS-INSU,\,Institut\,de\,Plan\'etologie\,et\,d'Astrophysique\,de\,Grenoble\,(IPAG),\,UMR\,5274,\,38041\,Grenoble,\,France
\and
Laboratoire\,d'Astrophysique\,de\,Marseille,\,Univ.\,de\,Provence,\,CNRS\,(UMR6110),\,38\,r.\,F.\,Joliot\,Curie,\,13388\,Marseille\,cedex\,13,\,France
\and
Centro de Astrof{\'\i}sica, Universidade do Porto, Rua das Estrelas, 4150-762 Porto, Portugal
\and
Observatoire de Gen\`eve,  Universit\'e de Gen\`eve, 51 Chemin des Maillettes, 1290 Sauverny, Switzerland
\and
Departamento de F\'{\i}sica e Astronomia, Faculdade de Ci\^encias, Universidade do Porto, Portugal
}

   \date{Received TBC; accepted TBC}
      
  \abstract{We observed with the \sophie\ spectrograph (OHP, France) 
  the transit of the \cibleb\ exoplanet across its host star. The resulting stellar
  radial velocities display the Rossiter-McLaughlin anomaly
  and reveal a retrograde orbit: 
  the planetary orbital spin and the stellar  rotational spin 
  point towards approximately opposite directions. 
  A~fit to the anomaly measures a sky-projected angle
  $\lambda = 166^{\circ} \pm 10^{\circ}$  between these two spin axes.
  All seven known retrograde planets are hot jupiters 
  with masses $M_p < 3\,$\MJ. About two thirds of 
  the planets in this mass range however are 
  prograde and aligned ($\lambda \simeq 0^{\circ}$). By 
  contrast, most of the more massive planets ($M_p > 4\,$\MJ) 
  are prograde but misaligned. 
  Different mechanisms may therefore be responsible
  for planetary obliquities above and below $\sim3.5\,$\MJ.
    }


  \authorrunning{H\'ebrard et al.}
\titlerunning{The retrograde orbit of the \cibleb\ exoplanet 
}

   \maketitle 


\section{Introduction}
\label{sect_introduction}

Spectroscopic observations during the transit of an exoplanet 
across its host star can measure the sky-projected
angle between the spins of the planetary orbit and
the stellar rotation (the obliquity) through 
the Rossiter-McLaughlin (RM) effect (Holt~\cite{holt93}; 
Rossiter~\cite{rossiter24}; McLaughlin~\cite{mclaughlin24}). 
The occultation of a rotating star by a planet distorts the 
apparent stellar line shape by removing the profile part
emitted by the hidden portion of the star. This induces 
anomalous stellar radial velocity variations during the transit,
which constrain the sky-projected obliquity ($\lambda$) 
and thus indicate whether the orbit is prograde, retrograde, or~polar. 

Queloz et al.~(\cite{queloz00}) reported the first detection of 
the RM anomaly for an extrasolar planet, HD\,209458b. That planet
shows an aligned, prograde orbit, as did all of the first
seven planets for which the RM effect was measured (as reviewed 
in H\'ebrard et al.~\cite{hebrard08}). These early results were 
interpreted as a validation of theories of planetary 
formation and evolution where a single giant planet migrates 
in a proto-planetary disk perpendicular to the 
stellar spin axis (e.g. Lin et al.~\cite{lin96}). 
That migration is expected to conserve the initial 
alignment between the angular momentums of the disk and of the 
planetary orbits.

H\'ebrard et al.~(\cite{hebrard08}) however found a first case 
of spin-orbit misalignment for the XO-3b planet, confirmed by 
Winn et al.~(\cite{winn09a}). Thereafter, Moutou et 
al.~(\cite{moutou09})
reported a second case, HD\,80606  (see also Pont et al.~\cite{pont09}; 
Winn et al.~\cite{winn09b}; H\'ebrard et al.~\cite{hebrard10}). 
A dozen misaligned systems 
have now been identified, including some with retrograde or 
nearly polar orbits (e.g. Winn et al.~\cite{winn09c}; Narita 
et al.~\cite{narita10a}; Triaud et al.~\cite{triaud10}; 
Simpson et al.~\cite{simpson10}). These unexpected results 
favor alternative scenarios where close-in massive planets
have been brought in by planet-planet (or planet-star) scattering, 
Kozai migration, and/or tidal friction (e.g. Malmberg et 
al.~\cite{malmberg07}; Fabrycky \& Tremaine~\cite{fabrycky07};
Chatterjee et al.~\cite{chatterjee08}; Nagasawa et al.~\cite{nagasawa08};
Guillochon et al.~\cite{guillochon10}).  Alternatively, it has been
proposed that the orbit still reflects the orientation of the disk,
with the stellar spin instead having moved away, either early-on 
through magnetosphere-disk interactions (Lai et al.~\cite{lai10}),
or later through elliptical tidal instability (C\'ebron et al.~\cite{cebron11}).
Distinguishing between these mechanisms
needs additional obliquity measurements (e.g. Morton 
\& Johnson~\cite{morton10}).

Here we present spectroscopic observations of one transit of \object{\cibleb}.
This hot jupiter transits a bright F star ($V=10.5$) every 3.8 days
and was discovered by Noyes et al.~(\cite{noyes08}, N08). Its mass is 
$1.06 \pm 0.12$~\MJ\ and its radius $1.33 \pm 0.06$~\RJ.

\section{Radial velocity measurements with \sophie}

The August 21$^{\mathrm{th}}$ 2010 transit of \cibleb\ was observed with 
\sophie. This cross-dispersed, stabilized 
echelle spectrograph is dedicated to high-precision radial velocity 
measurements (Perruchot et al.~\cite{perruchot08}; Bouchy et 
al.~\cite{bouchy09}). It is fed by two optical fibers mounted 
at the focus of the 1.93-m telescope of the Haute-Provence 
Observatory (OHP, France). \cible\ is bright enough for observation
in the high-resolution mode of the spectrograph 
($\lambda/\Delta\lambda=75,000$)  and with fast detector~readout.

The observations were carried out three days before full Moon. 
This was anticipated not to adversely impact the 
radial velocity accuracies, however, thanks to the large radial 
velocity shift between \cible\ (-22.7\,\kms, N08) and the Moon
(close to the barycentric Earth radial velocity, which in the 
direction of \cible\ was 16.3~\kms\ during the observations). We 
used the second fiber input to measure the sky background, and confirmed that 
its radial velocity was always shifted by at least 30\,\kms\ from that of \cible. 
We therefore did not subtract
the sky background as this would have introduced additional~noise.

We collected  41 measurements during the transit night 
(JD\,$\simeq2\,455\,430.5$) between 20:10 and 03:40 UT. 
The sky was clear and the seeing stable around $1\farcs6$. 
The airmass first decreased from sec$\,z=1.8$ to $1.0$ then 
increased to 1.1 at the end of the night.
We adjusted the exposure times between 500 and 800\,s to 
maintain a constant signal-to-noise ratio (SNR) of 39 per 
pixel at 550~nm. Spectra of a thorium lamp were obtained 
at the beginning and end of the 7.5-hour sequence, as well
as three times during the sequence. These wavelength calibration
exposures show that the spectrograph drifted by $\sim2\;$\ms\ 
per hour. We linearly interpolated between the calibration 
points to correct for this drift.

We used the \sophie\ pipeline (Bouchy et al.~\cite{bouchy09}) 
to extract the spectra, to cross-correlate them with a G2-type 
numerical mask, and to measure the radial velocities through
Gaussian fits to the cross-correlation functions (CCFs)
(Baranne et al.~\cite{baranne96}; Pepe et al.~\cite{pepe02}). 
We discarded the four bluest orders of the \sophie\ spectra that 
have low SNR; this slightly improves the dispersion 
of the data around the model. Every spectrum produces a clean CCF 
with a peak contrast of $11.45 \pm 0.10$\,\%\ of the continuum 
and a full width at half maximum
FWHM\,=\,$12.61 \pm 0.08$~km\,s$^{-1}$. The photon-noise 
uncertainty of the radial velocities, determined from the SNR 
of the spectra and the contrast and the FWHM
of the CCF (see Boisse et al.~\cite{boisse10} for the details),
is typically $\pm18\,$\ms. It is quite large by \sophie\ standard because 
the star is early-type and a moderately fast~rotator 
($V \sin i_s = 8.0 \pm 1.0$\,\kms\ and $T_{\rm eff} = 6570 \pm 80$\,K; 
see below).

Figure~\ref{fig_fit} shows the  \sophie\ velocities 
(see also the electronic table) as well as the 
orbital solution and HIRES data from N08. 
Outside transit, the \sophie\ data are compatible (within
the expected arbitrary offset) with the HIRES 
orbit. During transit (lower panel), the \sophie\ 
data have obvious residuals
relative to a Keplerian orbit, with a pattern opposite that
expected for an aligned, prograde transit ($\lambda=0^{\circ}$), 
red-shifted radial velocity in the first half of the transit
followed by a symmetric blue shift. Here, the observed pattern of 
blue-shifts during the first part of the transit and red-shifts 
in the second part, instead, is characteristic of the RM anomaly 
with a retrograde~orbit.

\begin{figure}[h] 
\begin{center}
\includegraphics[scale=0.57]{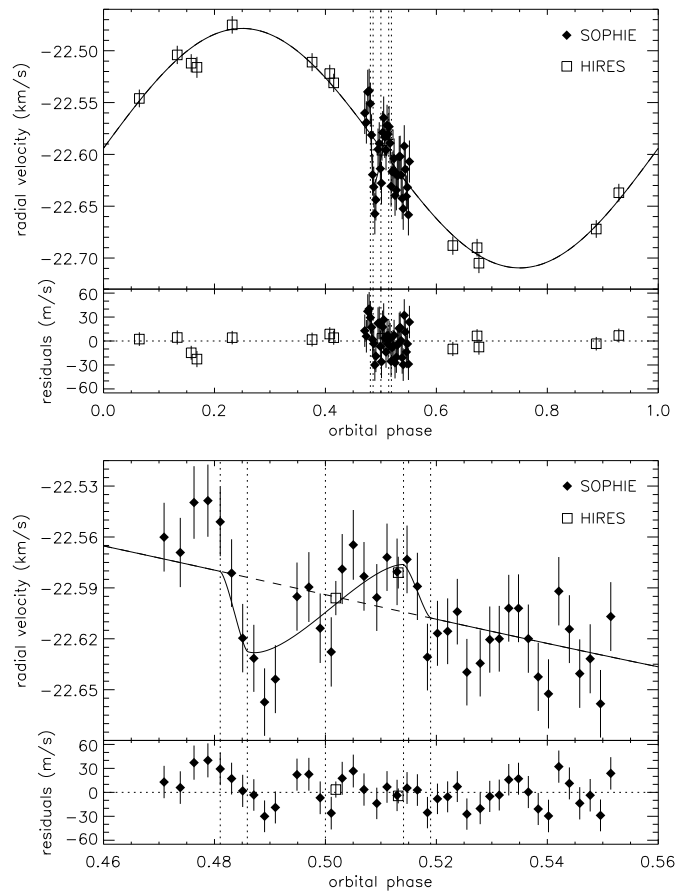}
\caption{Radial velocity measurements of \cible\ as a function of 
orbital phase. Filled diamonds represent the \sophie\ data (this paper) 
and open squares the HIRES data (N08). The lower panel zooms on 
the transit phases. The dashed line represents the Keplerian fit 
(ignoring the transit phases) while the solid line represents 
the final fit including the model of the Rossiter-McLaughlin anomaly. 
In each panel, the residuals of the final fit are plotted below the data.
The vertical dotted lines show the times of mid-transit, first, second, third, 
and fourth contacts.  The fitted values are
$V \sin i_s = 7.5 \pm 1.6$\,\kms\ and $\lambda = 166^{\circ} \pm 10^{\circ}$.
}
\label{fig_fit}
\end{center}
\vspace{-0.6cm}
\end{figure}

\section{Analysis}
\label{sect_analysis}

We fit our \sophie\ data together with the HIRES data from N08 to determine
the sky-projected obliquity 
in the \cible\ system. Interestingly, two HIRES 
measurements were serendipitously obtained in transits, 
on  December 16$^{\mathrm{th}}$, 2006 and July 4$^{\mathrm{th}}$,
2007 (the two squares in the lower panel of Fig.~\ref{fig_fit}). They
agree with the \sophie\ measurements secured at the same orbital 
phases and thus with a retrograde~orbit.

The orbital elements of N08 determine the center of 
the August 21$^{\mathrm{st}}$ 2010 transit, 
$T_\mathrm{tr} =2\,455\,430.4563 \pm 0.0018$\,BJD.
This $\pm2.6$-min uncertainty is negligible for our
analysis.~The~22 \sophie\ measurements secured 
outside transit determine the systemic, barycentric stellar 
velocity as $\gamma_{\mathrm{SOPHIE}} = -22.594 \pm 0.007$\,\kms.
The dispersion of their residuals from the orbital fit is 20.8\,\ms. 
The error bars in the electronic table and Fig.~\ref{fig_fit}
include an extra 9-\ms\ uncertainty added in quadrature to the photon noise  
to obtain a reduced \kid~of~1. Similarly the 13 HIRES measurements 
outside transit determine a systemic velocity  
$\gamma_{\mathrm{HIRES}} = -0.016 \pm 0.002$\,\kms\ (HIRES velocities are
relative) and have a 9.6-\ms\  dispersion around the Keplerian 
model. We quadratically added a 7.4-\ms\ jitter to their 
error bars (instead of the 8.6\,\ms\ adopted by N08) to obtain a \kid~of~1.
The \ion{Ca}{ii} activity level measured from the \sophie\ spectra,
$\log{R'_\mathrm{HK}}=-5.03 \pm 0.10$, agrees reasonably with 
$\log{R'_\mathrm{HK}}=-4.81$ measured by Wright~(\cite{wright05}).
It predicts a 10-\ms\  stellar jitter (Santos et 
al.~\cite{santos00}) consistent with these added dispersion~terms.

We model the RM anomaly with the analytical first-moment approach 
developed by Ohta et al.~(\cite{otha05}). We adopt the above values of the 
systemic velocities and mid-transit time, and the following  
parameters for the circular orbit taken from N08:
ratio of the planetary and
stellar radii $R_\mathrm{p}/R_* = 0.09338 \pm 0.00053$,
scaled semi-major axis $a / R_* = 7.69 \pm 0.22$, 
semi-amplitude of the radial velocity variation $K = 115.5 \pm 3.5$\,\ms, 
orbit inclination $i_o = 85\fdg51\pm0\fdg35$, and
orbital period $P = 3.852985 \pm 0.000005$~days.
We compute  from a model atmosphere (Kurucz~\cite{kurucz79}) 
a linear limb-darkening coefficient $\epsilon=0.57 \pm 0.10$ in 
the 5300-6300\,\AA\  wavelength range.

\begin{figure}[h] 
\begin{center}
\includegraphics[scale=0.51]{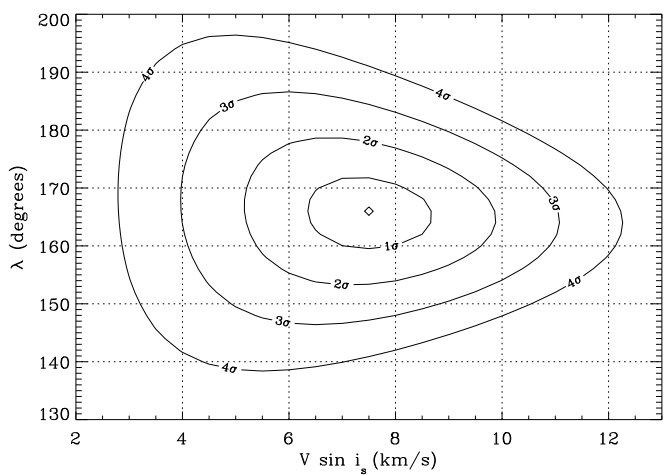}
\caption{\kid\ formal isocontours for our 
modeling of the Rossiter-McLaughlin effect as a function of $\lambda$ 
and  $V \sin i_s$ (see Sect.~\ref{sect_analysis}), using the formula from 
Ohta et al.~(\cite{otha05}) and the system parameters from N08. 
The diamond shows the best values.}
\label{fig_chi2}
\end{center}
\vspace{-0.2cm}
\end{figure}

The remaining two parameters of the Ohta et al. model are the sky-projected 
stellar rotational velocity $V \sin i_s$ and the sky-projected spin-orbit 
angle $\lambda$. We solve for these free parameters through a grid search,
and derive formal confidence intervals from the $\chi^2$ isocontours plotted 
in Fig.~\ref{fig_chi2} (H\'ebrard et al.~\cite{hebrard02}). The resulting 
values are  $V \sin i_s = 7.5 \pm 1.2$\,\kms\ and 
$\lambda = 166\fdg0 \pm 6\fdg1$. We separately propagate the effects 
of the uncertainties of the above fixed parameters on $V \sin i_s$ and $\lambda$,
and find a $\pm1.1$\,\kms\ effect on $V \sin i_s$, dominated by the 
uncertainties on $a / R_*$, $i$, and $\gamma_{\mathrm{SOPHIE}}$, and a 
$\pm8\fdg5$ effect on $\lambda$, mainly from the $\gamma_{\mathrm{SOPHIE}}$
uncertainty. The other fixed parameters contribute negligibly to 
the $\lambda$ and $V \sin i_s$ uncertainties. Quadratic addition of
the two sources of uncertainties produces our final results: 
$V \sin i_s = 7.5 \pm 1.6$\,\kms\ and $\lambda = 166^{\circ} \pm 10^{\circ}$.

The resulting $V \sin i_s$ agrees with the $V \sin i_s = 8.0 \pm 1.0$\,\kms\  
which we derive from the width of the CCF of  \cible\ (Boisse et 
al.~\cite{boisse10}), as well as with $V \sin i_s = 8.2 \pm 1.0$\,\kms\  
which N08 similarly derived from line broadening. The good agreement is
somewhat surprising, since rotation broadens the lines of \cible\ 
by significantly more than the \sophie\ resolution. As discussed
by e.g. Hirano et al.~(\cite{hirano10a}) and Simpson et al.~(\cite{simpson10}), 
naive modeling of the RM anomaly could produce biased $V \sin i_s$ measurements 
under such circumstances.

Figure~\ref{fig_fit} plots the full dataset and the final fitted model. The 
\sophie\ and HIRES velocities have dispersions around this model
of respectively 19.5\,\ms\ and 9.0\,\ms. These are 
similar to the dispersions of the radial velocities measured
outside transit around the Keplerian model, 20.8\,\ms\  and 
9.6\,\ms\ (see above). Our RM model is thus a good description of the 
measurements, with a reduced \kid\ of~unity. 

\begin{figure}[h] 
\hspace{-0.7cm}
\includegraphics[scale=0.34,angle=90]{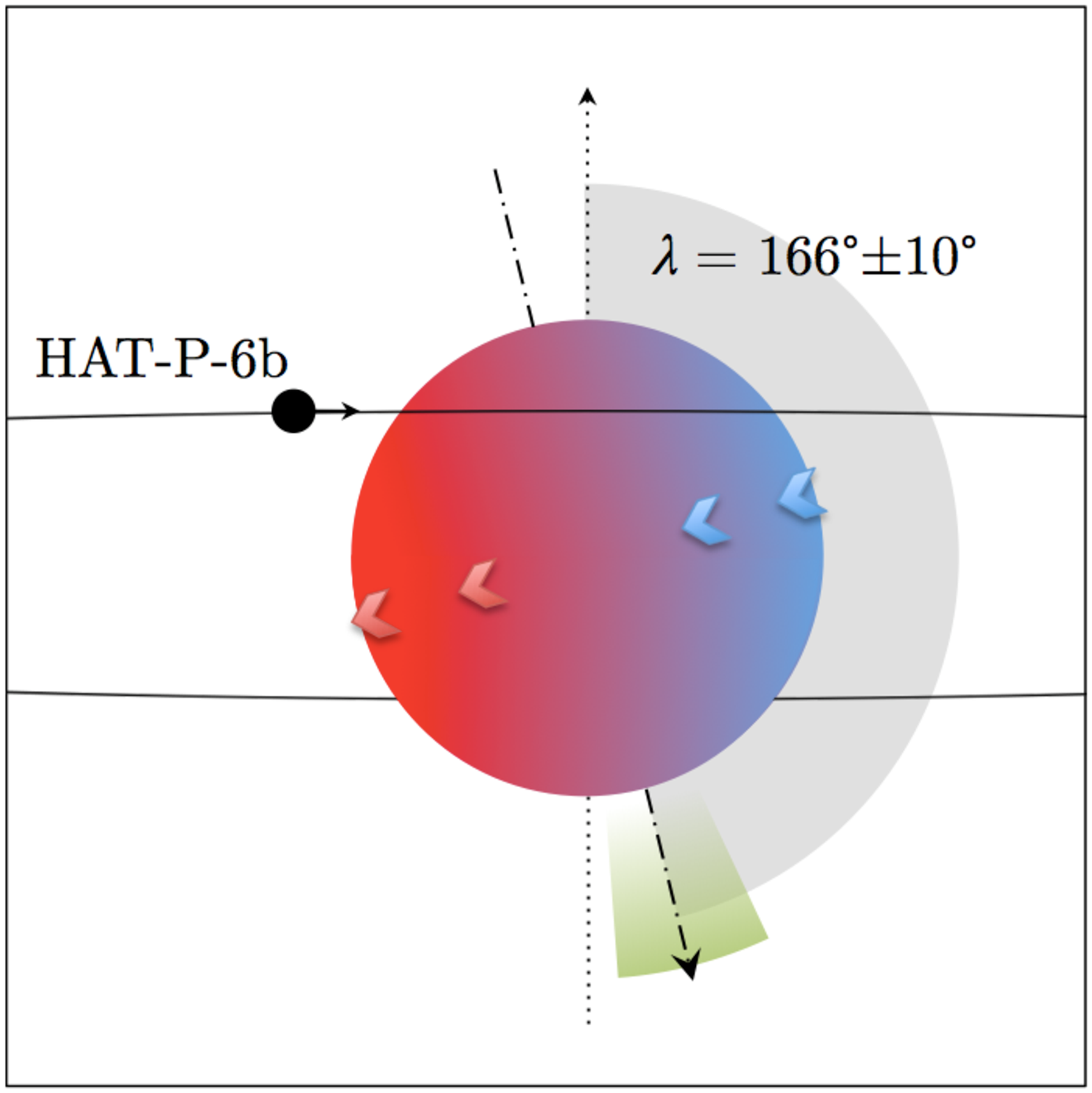}
\caption{Schematic view of the \cible\ system from the 
Earth. The dotted and dashed-dotted lines represent the orbital angular 
momentum and the sky-projection of the stellar 
spin axis. The red and blue hues on the star surface indicate the 
Doppler shift from stellar rotation. The sky-projected obliquity 
$\lambda$ and its uncertainty are displayed in grey and~green.}
\label{fig_orb}
\end{figure}

\section{Discussion}

The orbit of \cibleb\ is retrograde with respect to the spin of its host star
(Fig.~\ref{fig_orb}), but the true, unprojected
angle $\psi$ between the two angular momenta remains uncertain. 
We have only measured the \textit{sky-projected} spin-orbit angle, 
$\lambda = 166^{\circ} \pm 10^{\circ}$, related to $\psi$ by
$\cos\psi = \sin i_s \cos \lambda  \sin i_o  + \cos i_s \cos i_o $
(e.g. Fabrycky \& Winn~\cite{fabrycky09}, whose notations and definitions
we adopt:  $i_o$ is the inclination of the planetary orbit and  $i_s$ 
that of the stellar~spin). The angle $i_o$ is precisely known from the planetary
transits but $i_s$ remains unknown. The star cannot be {\it a priori} 
assumed to be seen edge-on ($i_s \simeq i_o \simeq 90^{\circ}$), 
especially here where $\lambda \neq 0$ demonstrates some misaligment.
Comparisons of the measured $V \sin i_s$ to estimate the 
stellar rotation period $P_\mathrm{rot}$ and radius  $R_*$ can
in principle constrain $i_s$, but here with ambiguous~results.

On one hand, the measured $R'_\mathrm{HK}$ implies a short stellar 
rotation period, $P_\mathrm{rot}\simeq3.5$~days (Mamajek \& 
Hillenbrand~\cite{mamajek08}). From  $R_*  = 1.46 \pm 0.06$~\RS\ 
and $V \sin i_s = 8.0 \pm 1.0$\,\kms, we obtain 
$P_\mathrm{rot} / \sin i_s = 9.2 \pm 1.5$~days. Reconciling
the two numbers requires an almost pole-on star, with 
$i_s \simeq 20^{\circ}$ or $160^{\circ}$\footnote{The
$i_s \simeq 200^{\circ}$ and $340^{\circ}$ formal solutions 
can be excluded since we know that the orbit is retrograde 
($|\lambda| > \pi - i_o$).}, implying $\psi \simeq 110^{\circ}$.
Schlaufman~(\cite{schlaufman10}), on the other hand, predicts 
$V_\mathrm{rot} = 9.1$\,\kms\ 
from the age and mass of \cible.
This is similar to the measured $V \sin i_s$, so would instead
imply a more edge-on star, with $i_s \simeq  60^{\circ}$ or 
$120^{\circ}$\footnote{Here again, the retrograde orbit excludes 
$i_s \simeq  240^{\circ}$ and $300^{\circ}$.}, 
and $\psi \simeq 145^{\circ}$. Neither approach is expected to be
accurate. Here they give inconsistent estimates for $P_\mathrm{rot}$
and therefore for $i_s$ and $\psi$. Asteroseismology 
(Gizon \& Solanki~\cite{gizon03}), 
polarization of magnetic 
dipoles  (L\'opez-Ariste et al.~\cite{lopez10}), or accurate photometry
may hopefully provide reliable $i_s$ and $P_\mathrm{rot}$ measurements.
In the mean time the value of  $\psi$ remains unknown, except for
its being above $90^{\circ}$.

\cibleb\ is the seventh planet identified as having a retrograde orbit, 
after WASP-2b, 8b, 15b and 17b (Triaud et al.~\cite{triaud10}; 
Queloz et al.~\cite{queloz10}; Bayliss et al.~\cite{bayliss10})
and HAT-P-7b and 14b (Winn et al.~\cite{winn09c}, \cite{winn10a};
Narita et al.~\cite{narita09}). Three planets additionally seem
to have nearly polar orbits: CoRoT-1b (Pont et al.~\cite{pont10}), 
WASP-1b (Simpson et al.~\cite{simpson10}), and 
HAT-P-11b (Winn et al.~\cite{winn10b};  Hirano~\cite{hirano10b}).
The last of those is also likely retrograde. HAT-P-11b in addition
is the only Neptune-mass planet with an obliquity measurement,
with all other planetary systems with RM measurements being 
jovian mass planets.

The processes that produce tilted systems remain a matter of debate
(see Sect.~\ref{sect_introduction}). Whether 
a single mechanism explains all close-in planets or whether 
different processes produce aligned and misaligned systems is
unknown. Even whether the planetary orbits acquired 
obliquity or the stellar spin instead acquired tilt is uncertain. 
Those mechanisms, in any case, cannot require too narrowly specified
conditions, since tilted systems are common.

Schlaufman~(\cite{schlaufman10}) and Winn et al.~(\cite{winn10c}) 
hypothesized that misaligned planets preferentially orbit
hot stars. The $6570 \pm 80$\,K effective temperature of \cible\  
(N08) and the retrograde orbit of \cibleb\ support this trend.
As suggested by Johnson et al.~(\cite{johnson09}) and H\'ebrard et 
al.~(\cite{hebrard10}), 
planetary mass apparently is also a key parameter. The 
seven retrograde and three polar planets all have masses below 
3\,\MJ. In that mass range, about a third of the 
planets with RM observations presents such extremes obliquities, 
with the other two thirds having $\lambda \simeq 0^{\circ}$. 

Measured planets with masses above 4\,\MJ\ are fewer but they 
seem to have a different obliquity distribution: 
four (over six) of these massive planets are misaligned,
but none with ultra-high obliquities, i.e. none on retrograde orbit and 
even none with $|\lambda| > 50^{\circ}$. Just two of the massive planets show 
$\lambda \simeq 0^{\circ}$, HAT-P-2b (Winn et al.~\cite{winn07}; 
Loeillet et al.~\cite{loeillet08}) and WASP-18 (Triaud et 
al.~\cite{triaud10}). Curiously, the period distribution of 
exoplanets detected from radial velocity surveys changes
at a similar characteristic mass. There is a distinct lack of 
planets with $M_p>4$\,\MJ\ in short-period orbits 
($< 100$\, days), and radial velocity surveys detect no 
planets above that limit outside multiple star systems
(Udry et al.~\cite{udry03}; Boisse et al.~\cite{boisse10}). 

\begin{figure}[h] 
\begin{center}
\includegraphics[scale=0.51]{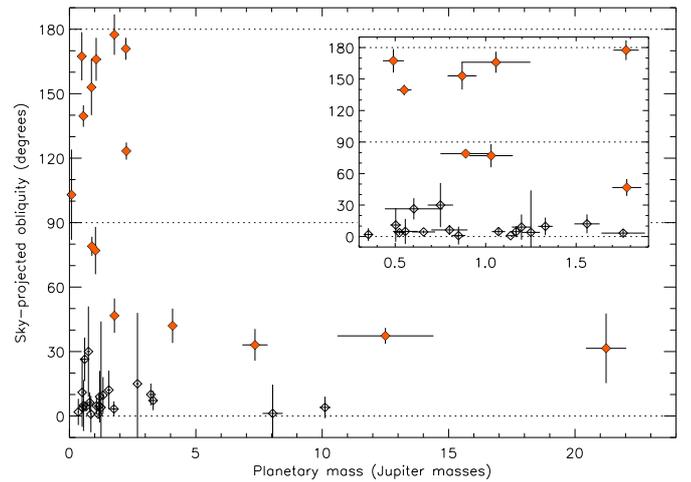}
\caption{Absolute value of $\lambda$, the sky-projected obliquity, 
as a function of planet mass $M_p$ for the 37 planets with 
published RM measurements. In-filled, red markers denote
systems with $|\lambda | > 30^{\circ}$. The inset zooms on 
$0.3$\,\MJ$\,<M_p < 1.9\,$\MJ. The dotted lines mark
aligned-prograde, polar, and aligned-retrograde systems
($|\lambda | = 0^{\circ}, 90^{\circ}$, and $180^{\circ}$, respectively).
The 37 plotted planetary systems are the 21 ones compiled in H\'ebrard 
et al.~(\cite{hebrard10}), plus XO-4 (Narita et al.~\cite{narita10a}), 
WASP-1, 24, and 38 (Simpson et al.~\cite{simpson10}),
WASP-2, 4, 5, 15, and 18 (Triaud et al.~\cite{triaud10}),
WASP-8 (Queloz et al.~\cite{queloz10}),
WASP-19 (Hellier et al.~\cite{hellier11}),
HAT-P-4 and 14 (Winn et al.~\cite{winn10a}),
HAT-P-8 (Simpson et al.~\cite{simpson10}),
HAT-P-11 (Winn et al.~\cite{winn10b}; Hirano~\cite{hirano10b}), and
HAT-P-6 (this work).
}
\label{fig_revue}
\end{center}
\vspace{-0.5cm}
\end{figure}

The changes in the $\lambda$-distribution with planetary mass 
is illustrated in Fig.~\ref{fig_revue} 
by the 37 systems with published $\lambda$ and $M_p$. The inset 
zooms on the Jupiter-mass planets, with their 2/3 of aligned and 
1/3 strongly misaligned planets. The massive planets, by contrast,
mostly show significant but moderate misalignments. This suggests 
that distinct mechanisms dominate for different planet masses, 
with a critical mass near $3.5\,$\MJ.
If a combination of planet scattering, Kozai mechanism, and
tidal circularization explains most oblique, close-in planets 
(Nagasawa et al.~\cite{nagasawa08}; Morton \&
Johnson~\cite{morton10}), planetary mass is a natural control
parameter. These conclusions however currently suffer from 
small-number statistics. Additional Rossiter-McLaughlin observations 
of massive planets are needed to confirm 
that most of them are tilted and none of them are polar nor~retrograde.

\begin{acknowledgements}
We thank the Haute-Provence Observatory staff who support the 
operation of \sophie.
We acknowledge support of the ``Programme National de Plan\'etologie'' 
(PNP) of CNRS/INSU, 
the Swiss National Science Foundation, 
and the French National Research Agency (ANR-08-JCJC-0102-01 and 
ANR-NT05-4-44463).
D.E. is supported by CNES. 
A.E. is supported by a fellowship for advanced researchers from the 
Swiss National Science Foundation.
I.B. and N.C.S. acknowledge the support by the European Research 
Council/European Community under 
the FP7 through Starting Grant agreement number 239953, as well as 
the support from Funda\c{c}\~ao 
para a Ci\^encia e a Tecnologia (FCT) through program Ci\^encia\,2007 
funded by FCT/MCTES
(Portugal) and POPH/FSE (EC), and in the form of grants reference 
PTDC/CTE-AST/66643/2006 
and PTDC/CTE-AST/098528/2008.
\end{acknowledgements}


\begin{thebibliography}{}

\bibitem[1996]{baranne96} 
Baranne, A., Queloz, D., Mayor, M., et al. 1994, \aaps, 119, 373

\bibitem[2010]{bayliss10} 
Bayliss, D., Winn, J., Mardling, R., Sackett, P. 2010, \apj, 722, L224

\bibitem[2010]{boisse10} 
Boisse, I., Eggenberger, A., Santos, N. C., et al. 2010, \aap, 523, A88

\bibitem[2009]{bouchy09} 
Bouchy, F., H\'ebrard, G., Udry, S., et al. 2009, \aap, 505, 853

\bibitem[2011]{cebron11} 
C\'ebron,\,D.,\,Moutou,\,C.,\,Le\,Bars,\,M.,\,Le\,Gal,\,P.,\,Fares,\,R.\,2011\,[arXiv:1101.4531]

\bibitem[2008]{chatterjee08} 
Chatterjee, S., Ford, E. B., Matsumura, S., 
Rasio, F. A. 2008, \apj, 686, 580

\bibitem[2007]{fabrycky07}
Fabrycky, D., Tremaine, S. 2007, \apj, 669, 1298

\bibitem[2009]{fabrycky09}
Fabrycky, D. C., Winn, J. N. 2009, \apj, 696, 1230

\bibitem[2003]{gizon03}
Gizon, L., Solanki, S. K. 2003, \apj, 589, 1009

\bibitem[2010]{guillochon10}
Guillochon, J., Ramirez-Ruiz, E., Lin, D. N. 2010,  \apj, sub. [arXiv:1012.2382]

\bibitem[2002]{hebrard02} 
H\'ebrard, G., Lemoine, M., Vidal-Madjar, A., et al. 2002, \apj, 140, 103

\bibitem[2008]{hebrard08} 
H\'ebrard, G., Bouchy, F., Pont, F., et al. 2008, \aap, 481, 52

\bibitem[2010]{hebrard10} 
H\'ebrard, G., D\'essert, J.-M, D\'{\i}az, R.~F., et al. 2010, \aap, 516, A95

\bibitem[2011]{hellier11} 
Hellier, C., Anderson, D. R., Collier Cameron, A., et al. 2011 [arXiv:1009.5677]

\bibitem[2010a]{hirano10a} 
Hirano, T., Suto, Y., Taruya, A., et al.
2010a, \apj, 709, 458

\bibitem[2010b]{hirano10b} 
Hirano, T., Narita, N., Shporer, A., et al. 
2010b, \pasj, sub. [arXiv:1009.5677]

\bibitem[1893]{holt93} 
Holt, J. R., 1893, Astronomy and Astrophysics, XII

\bibitem[2009]{johnson09} 
Johnson, J. A., Winn, J. N., Albrecht, S., et al.
2009, \pasp, 121, 1104

\bibitem[1979]{kurucz79} 
Kurucz, R. L. 1979, \apjs, 40, 1

\bibitem[2010]{lai10} 
Lai, D., Foucart, F., Lin,  D. N. C. 2010, \mnras, in press [arXiv:1008.3148]

\bibitem[1996]{lin96} 
Lin, D. N. C., Bodenheimer, P., Richardson, D. C. 1996, \nat, 380, 606

\bibitem[2008]{loeillet08} 
Loeillet, B., Shporer, A., Bouchy, F., et al. 2008, \aap, 481, 529

\bibitem[2010]{lopez10}
L\'opez-Ariste, A., 
et al. 
2010, \aap, in press [arXiv:1011.6288]

\bibitem[2007]{malmberg07} 
Malmberg, D., Davies, M. B., Chambers, J. E. 2007, \mnras, 377, L1

\bibitem[2008]{mamajek08} 
Mamajek, E. E., \& Hillenbrand, L. A. 2008, \apj, 687, 1264

\bibitem[1924]{mclaughlin24} 
McLaughlin, D. B., 1924, \apj, 60, 22

\bibitem[2010]{morton10}
Morton, T. D., Johnson, J. A. 2010, \apj, in press [arXiv1010.4025]

\bibitem[2009]{moutou09}
Moutou, C., H\'ebrard, G., Bouchy, F., et al. 2009, \aap, 498, L5 

\bibitem[2008]{nagasawa08}
Nagasawa, M., Ida, S., Bessho, T. 2008, \apj, 678, 498

\bibitem[2009]{narita09}
Narita, N., Sato, B., Hirano, T., Tamura, M. 2009, \pasj, 61, L35

\bibitem[2010a]{narita10a}
Narita, N., Hirano, T., Sanchis, R., et al.
2010a, \pasj, 62, L61

\bibitem[2008]{noyes08}
Noyes, R. W.,  Bakos, G. \`A, Torres, G., et al. 2008, \apj, 673, L79 (N08)

\bibitem[2005]{otha05} 
Ohta, Y., Taruya, A, Suto, Y. 2005, \apj, 622, 1118

\bibitem[2002]{pepe02} 
Pepe, F., Mayor, M., Galland, F., et al. 2002, \aap, 388, 632

\bibitem[2008]{perruchot08}
Perruchot, S., Kohler, D., Bouchy, F., et al. 2008, 
SPIE proceedings, 70140J

\bibitem[2009]{pont09}
Pont, F., H\'ebrard, G., Irwin, J. M., et al. 2009, \aap, 509, 695

\bibitem[2010]{pont10}
Pont, F., Endl, M., Cochran, W. D., et al. 2010, \mnras, 402, L1 

\bibitem[2000]{queloz00}
Queloz, D., Eggenberger, A., Mayor, M., et al.
2000, \aap, 359, L13

\bibitem[2010]{queloz10}
Queloz, D., Anderson, D., Collier Cameron, A., et al. 2010, \aap, 517, L1

\bibitem[1924]{rossiter24} 
Rossiter, R. A., 1924, \apj, 60, 15

\bibitem[2000]{santos00} 
Santos, N. C., Mayor, M., Naef, D., et al. 2000,  \aap, 361, 265

\bibitem[2010]{schlaufman10}
Schlaufman, K. C. 2010, \apj, 719, 602

\bibitem[2010]{simpson10} 	
Simpson, E. K., 
et al. 2010, \mnras, sub.  [arXiv:1011.5664]

\bibitem[2010]{triaud10} 	
Triaud, A., Collier Cameron, A., Queloz, D., et al. 2010, \aap, 524, A25

\bibitem[2003]{udry03} 	
Udry, S., Mayor, M., Santos, N. C. 2003, \aap, 407, 369

\bibitem[2007]{winn07} 	
Winn, J. N., Johnson, J. A., Peek, K. M. G., et al. 2007, \apj, 665, L167

\bibitem[2009a]{winn09a} 	
Winn, J. N., Johnson, J. A., Fabrycky, D., et al. 2009a, \apj, 700, 302	

\bibitem[2009b]{winn09b} 	
Winn, J. N., Howard, A. W., Johnson, J. A., et al. 2009b, \apj, 703, 2091

\bibitem[2009c]{winn09c} 	
Winn, J. N., Johnson, J. A., Albrecht, S., et al.
2009c, \apj, 703, L99

\bibitem[2010a]{winn10a} 	
Winn, J. N., Howard, A., Johnson, J., et al. 2010a, \aj, 141, 63

\bibitem[2010b]{winn10b} 	
Winn, J. N., Johnson, J. A., Howard, A. W., et al. 2010b, \apj, 723, L223

\bibitem[2010c]{winn10c} 	
Winn, J. N., Fabrycky, D., Albrecht, S., Johnson, J. A. 2010c, \apj, 718, L145

\bibitem[2005]{wright05} 	
Wright, J. T. 2005, \pasp, 117, 657

\end{thebibliography}
\end{document}